\documentclass[twocolumn]{aastex7}
\usepackage{graphicx}
\usepackage{subfigure}
\usepackage{color}
\usepackage{amsmath}
\usepackage{soul}

\shorttitle{Radiation Spectrum of a Turbulent Photosphere}
\shortauthors{Li et al.}
\begin{document}
\title{Radiation Spectrum of the Photospheric Emission for a Turbulent Relativistic Jet}

\author[orcid=0000-0002-6727-7798, gname=Guo-Yu, sname=Li]{Guo-Yu Li}
\affiliation{Guangxi Key Laboratory for Relativistic Astrophysics, School of Physical Science and Technology, Guangxi University, Nanning 530004, China}
\email{guoyu.li@st.gxu.edu.cn}

\author[orcid=0000-0003-1474-293X, gname=Da-Bin, sname=Lin]{Da-Bin Lin}
\affiliation{Guangxi Key Laboratory for Relativistic Astrophysics, School of Physical Science and Technology, Guangxi University, Nanning 530004, China}
\email[show]{lindabin@gxu.edu.cn}
\correspondingauthor{Da-Bin Lin}

\author[orcid=0000-0002-0926-5406, gname=Zhi-Lin, sname=Chen]{Zhi-Lin Chen}
\affiliation{Guangxi Key Laboratory for Relativistic Astrophysics, School of Physical Science and Technology, Guangxi University, Nanning 530004, China}
\email{2007301015@st.gxu.edu.cn}

\author[orcid=0000-0002-7044-733X, gname=En-Wei, sname=Liang]{En-Wei Liang}
\affiliation{Guangxi Key Laboratory for Relativistic Astrophysics, School of Physical Science and Technology, Guangxi University, Nanning 530004, China}
\email{lew@gxu.edu.cn}
\begin{abstract}
The prompt $\gamma$-rays of gamma-ray bursts (GRBs) may originate from the photosphere of a relativistic jet.
However,
only a few GRBs have been observed with evident blackbody-like emission,
for example, GRB~090902B.
It has been demonstrated that internal dissipation processes,
such as magnetic reconnection,
can occur within the relativistic jet
and thereby drive violent turbulence in the dissipation region.
In this paper,
we study the photospheric emission of a jet with turbulence below its photosphere.
Here, the turbulence is modeled phenomenologically
under the assumption that the four-velocity of its eddies follows a Gaussian distribution in the jet's co-moving frame.
It is found that the turbulence scatters photons to high energies and thus intensifies the emission in the high-energy regime.
The corresponding distortion of the radiation field can be preserved if and only if the turbulence occurs in a region with incomplete photon-electron coupling.
Consequently, the observed radiation spectrum can be reshaped into a Band-like spectrum.
\end{abstract}
\keywords{Gamma-ray bursts (629): general - turbulence - magnetic reconnection}
%%%%%%%%%%%%%%%%%%%%%%%%%%%%%%%%%
%%%%%%%%%%%%%%%%%%%%%%%%%%%%%%%%%
	
\section{Introduction}\label{Sec:Intro}
Gamma-ray bursts (GRBs),
first detected serendipitously in the late 1960s (\citealp{Klebesadel-1973ApJ...182L..85K}),
represent the most luminous electromagnetic explosions in the Universe (\citealp{2015PhR...561....1K}).
Despite extensive observational efforts, the physical origin of GRB prompt emission remains a major unsolved problem in high-energy astrophysics.
A key observational clue lies in their spectral properties: most GRB spectra are well-described by the empirical Band function (\citealp{Band-1993ApJ...413..281B}),
characterized by a smoothly connected broken power-law shape with low-energy photon spectral index $\alpha \sim -0.8$,
the high-energy photon spectral index $\beta \sim -2.3$,
and the break energy $E_{\rm 0} \sim 150\,{\rm keV}$ (\citealp{Yu-2016A&A...588A.135Y}).
However,
it is difficult to associate these parameters with the physical processes responsible for GRB emissions.
Two principal mechanisms dominate theoretical discussions:
synchrotron radiation (\citealp{1994ApJ...432..181M,1998MNRAS.296..275D,2000ApJ...543..722L,2011A&A...526A.110D,2011ApJ...726...90Z}) and photospheric emission (\citealp{1994MNRAS.270..480T,2000ApJ...530..292M,Beloborodov-2010MNRAS.407.1033B,2010ApJ...725.1137L}).

Synchrotron radiation plays a significant role in shaping GRB spectra (\citealp{2020NatAs...4..210Z}) but faces a critical challenge in its predicted low-energy spectral index.
The typical dissipation processes in the jets of GRBs usually predict that the electrons are in a fast cooling state,
leading to synchrotron emission with a low-energy photon spectral index $\alpha=-1.5$ (\citealp{1998ApJ...497L..17S,2000MNRAS.313L...1G}).
The low-energy photon spectral index predicted for synchrotron radiation is softer than observations of typical GRBs.
Moreover,
some GRBs have been observed to exhibit $\alpha > -1/3$,
which is beyond the ``death-line" of synchrotron radiation (\citealp{2018ApJ...864L..16Y}).
In contrast, the photospheric emission model successfully explains certain observational correlations,
such as the Amati relation and spectral peak-energy evolution (\citealp{2013ApJ...765..103L,2014MNRAS.442.2202L,2019NatCo..10.1504I}).
However,
its predicted spectrum is narrower than those of typical GRBs.
In the Rayleigh-Jeans regime,
pure blackbody emission is predicted to be with $\alpha \sim +1$,
but relativistic and geometric effects can soften the radiation spectrum to $\alpha \sim +0.4$
(\citealp{Beloborodov-2010MNRAS.407.1033B,Deng-2014ApJ...785..112D}).
Although narrow spectra have been observed in some GRBs like GRB~090902B (\citealp{2010ApJ...709L.172R,2011ApJ...730..141Z}) and GRB~220426A (\citealp{2022ApJ...934L..22D,2022ApJ...940..142W}),
most GRBs lack such signatures.

If dissipation occurs below the photosphere,
sub-photospheric electrons are heated through dissipation processes,
and Comptonization of photons affects the spectrum shape (\citealp{Rees-2005ApJ...628..847R,Pe'er-2006ApJ...652..482P}).
Simulations suggest that sub-photospheric dissipation can lead to a non-thermal radiation spectrum (\citealp{Ryde-2011MNRAS.415.3693R}).
Various dissipation mechanisms have been considered,
including magnetic reconnection (\citealp{Giannios-2005A&A...430....1G}),
internal shocks (\citealp{Rees-2005ApJ...628..847R}),
and collisional heating (\citealp{Beloborodov-2010MNRAS.407.1033B,Vurm-2011ApJ...738...77V}).
Additionally,
geometric effects can broaden the photospheric emission spectrum (\citealp{Pe'er-2008ApJ...682..463P,Deng-2014ApJ...785..112D,Meng-2018ApJ...860...72M}).
Despite extensive studies,
it remains challenging to obtain spectra with $\alpha \sim -0.8$ for the photospheric emission (\citealp{2019pgrb.book.....Z}).
We note that any dissipation mechanism in the jet inevitably drives turbulence.
This turbulence,
a common yet frequently overlooked feature of relativistic jets,
likely plays a key role in bridging the gap between theoretical predictions and observations.

The paper is organized as follows:
Section \ref{Sec:Model} provides a detailed explanation of our numerical model and describes how we incorporate isotropic turbulence and magnetic reconnection.
Section \ref{Sec:Result} presents simulation results for the photospheric emission of a turbulent jet.
Section \ref{Sec:Discussion} summarizes our results.

\section{Model}\label{Sec:Model}
\subsection{The MCRaT Code}
We use the Monte Carlo Radiation Transfer (MCRaT) code (\citealp{Lazzati-2016ApJ...829...76L,Parsotan-2018ApJ...853....8P,Parsotan-2018ApJ...869..103P,Parsotan-2020ApJ...896..139P}) to simulate photon propagation and scatter in a turbulent jet.
In the MCRaT code,
photons are injected within a radius range of $[R_{\rm inj} - \Delta{R},\, R_{\rm inj} + \Delta{R}]$.
In this paper, $\Delta{R} = R / (2 \Gamma_{\rm b})$ is set,
where $\Gamma_{\rm b}$ is the bulk Lorentz factor of the jet fluid and $R$ represents the jet radius measured relative to the central engine of the burst.

The expected number of photons in the $i$-th fluid element is given by
\begin{equation}\label{Eq.inject-number}
    \mathrm{d}{N_{i}} = \frac{\zeta T_{i}^{\prime 3} \Gamma_{i}}{\omega} \mathrm{d}{V_{i}},
\end{equation}
where the prime represents the parameters estimated in the co-moving frame of the jet,
$T_{i}^{\prime}$ is the temperature of the fluid element in a co-moving frame,
$\zeta$ is the number density coefficient ($\zeta=20.29$ for a Planck spectrum, $\zeta=8.44$ for a Wien spectrum),
$\Gamma_{i}$ is the Lorentz factor of the fluid element,
$\mathrm{d}{V_{i}}$ is the volume of the fluid element,
and $\omega$ is the weight factor of the injected photons.
The code dynamically adjusts the photon weight factor $\omega$ to match the predefined total number of photons in the injection region.
Each injected photon is assigned an azimuth angle $\phi'\in[0,2\pi]$ and a polar angle $\theta'\in[0,\pi]$, 
where $\phi'$ is randomly sampled from $[0, 2\pi]$ and $\cos\theta'$ is randomly sampled from $[-1,1]$.
The four-momentum of the $i$-th injected photon in the co-moving frame is defined as
\begin{equation}\label{photon-momentum}
p^{\prime}_{i}=\frac{h \nu^{\prime}_{i}}{c}(1;\,\sin\theta'_{i}\cos\phi'_{i};\,\sin\theta'_{i}\sin\phi'_{i};\, \cos\theta'_{i}),
\end{equation}
where $\nu^{\prime}_{i}$ is the frequency of the injected photons in the jet co-moving frame
and $h$ is the Planck constant.

Based on the density $\rho_i$ of the fluid element encompassing
the $i$-th photon,
one can obtain the mean free path for the corresponding photon (\citealp{1991ApJ...369..175A}), i.e.,
\begin{equation}
    \lambda_{i} = \frac{m_{\rm p}}{\rho_{i}\sigma_{\rm T} (1 - \beta_{i} \cos{\theta_{{\rm fp},i}})},
\end{equation}
where $\sigma_{\rm T}$ is the Thomson scattering cross-section,
$m_{\rm p}$ is the proton mass,
$\beta_{i}$ denotes the fluid velocity (in units of $c$), 
and $\theta_{{\rm fp},i}$ represents the angle between the fluid velocity vector and photon momentum.
Statistically, the scattering path length $l$ for a photon obeys the distribution of $f(l/{\lambda_{i}}) =\mathrm{e}^{-l/\lambda_{i}}$.
Consequently, the scattering path length for the $i$-th photon in MCRaT code is sampled as 
\begin{equation}
    l_{i} = - \frac{m_{\rm p}}{\rho_{i} \sigma_{\rm T} (1 - \beta_{i} \cos{\theta_{{\rm fp},i}})} \ln{\xi},
\end{equation}
where $\xi$ is a random number uniformly selected from [0, 1].
The scattering time is thus $t_{{\rm s},i} = l_{i}/c$.
The photon with the smallest $t_{{\rm s},i}$ is assumed to scatter first,
and thus the positions of all the photons are advanced by the
time of $t_{\rm s,min}=\min\{t_{{\rm s},i}\}$.
Although the photon with the smallest $t_{{\rm s},i}$ is the photon of interest to scatter,
the scattering is determined by the Klein-Nishina cross-section ($\sigma_{\rm KN}$) (\citealp{doi:10.13182/NSE11-57}),
i.e.,
if $\xi < \sigma_{\rm KN} / \sigma_{\rm T}$,
the photon scatters, updating its momentum and position;
otherwise, only its position is updated.
Here, $\xi$ is a random number uniformly selected from [0, 1].
For more details on the MCRaT code,
please refer to these articles, e.g.,
\cite{Lazzati-2016ApJ...829...76L,Parsotan-2018ApJ...853....8P,Parsotan-2018ApJ...869..103P,Parsotan-2020ApJ...896..139P}.

\subsection{Prescription of a Relativistic Jet}
\emph{Bulk motion of the jet\;\;}
We employ numerical modeling to determine the fluid properties of the jet,
including the bulk Lorentz factor ($\Gamma_{\rm b}$),
the co-moving temperature ($T^{\prime}$),
and the co-moving density ($\rho^{\prime}$).
The bulk Lorentz factor is related to the radius $R$ of the fluid with respect to the central engine, i.e.,
\begin{equation}\label{Eq.fireball-gamma}
\Gamma_{\rm b}(R)=\left\{ \begin{array}{cc}
{R}/{R_{0}}, &  R \leq R_{\rm s}\\
\eta, &  R > R_{\rm s}
\end{array} ,\right.
\end{equation}
where $R_{0}$ is initial radius of fireball,
$R_{\rm s}=\eta R_{0}$ is the saturation radius,
and $\eta$ is the dimensionless entropy of the jet.
The co-moving temperature is given by
\begin{equation}\label{Eq.fireball-T}
T^{\prime}_{\rm b}(R)=\frac{T_{0}}{2\Gamma_{\rm b}(R)}\times\left\{ \begin{array}{cc}
1, & R \leq R_{\rm s}\\
({R}/{R_{\rm s}})^{-\frac{2}{3}}, & R > R_{\rm s}
\end{array} ,\right.
\end{equation}
where $T_{0}=[L_{\rm j}/(4\pi R_{0}^{2}ac)]^{1/4}$ is the initial temperature for the fireball,
$a$ is the radiation density constant,
and $L_{\rm j}$ is the jet's power.
The co-moving density is expressed as
\begin{eqnarray}\label{Eq.fireball-rho}
	\rho^{\prime}=L_{\rm j}/(4 \pi R^{2} c^{3} \eta \Gamma_{\rm b}).
\end{eqnarray}

\emph{Turbulent motion of the fluid elements\;\;}
In addition to bulk motion described previously, 
turbulent motion also acts upon individual fluid elements.
Turbulence in astrophysical jets may be powered by diverse physical mechanisms.
Given the complexity of its origin,
we employ a phenomenological approach to model turbulent properties.
Our phenomenological model assumes isotropic turbulence in the co-moving frame of the jet,
consistent with the results in numerical simulations which show a Gaussian distribution of the turbulent velocity approximately (\citealp{1991FlDyR...8...65Y,2012ApJ...751...26E}).
Accordingly, 
we model the probability density functions for the three spatial components ($x', y', z'$) of the four-velocity field as Gaussian distributions.

Owing to the expansion of the jet,
we define a series of turbulent cells in the co-moving frame of the jet.
Each turbulent cell has a characteristic scale of $\lambda' = R/(\chi \Gamma_{\rm b})$,
where $\chi$ is a scale factor.
The corresponding active timescale for eddies is $\Delta t_{\rm turb} = \lambda' / c$.
We assume that turbulent eddies are uniformly distributed in space,
with a probability $p$ to exhibit turbulent activity for any eddy.
The turbulent four-velocity of the $i$-th turbulent cell,
denoted as ${\gamma'_{i} \beta'_{i,k}} |_{k \in \{x', y', z'\}}$,
is set as follows:
\begin{itemize}
\item if $\xi_i \leq p$,
the value of ${{\gamma'_{i} \beta'_{i,k}} |_{k \in \{x', y', z'\}}}$ is randomly selected based on the distribution function $\mathcal{N}(0, \gamma'^{2}_{0} \beta'^{2}_{0})$,
\item if $\xi_i > p$,
${{\gamma'_{i} \beta'_{i,k}} |_{k \in \{x', y', z'\}}}=0$ is set,
\end{itemize}
where $\xi_i$ is a number randomly selected in $[0, 1]$
and used to control the activity of the $i$-th cell,
$\beta'_{0} = \sqrt{\sigma/(1 + \sigma)}$, $\gamma'_{0}=({1 - {\beta'_{0}}^{2}})^{-1/2}$,
and $\mathcal{N}(0, \gamma'^{2}_{0} \beta'^{2}_{0})$ is a Gaussian distribution with mean zero and variance $\gamma'^{2}_{0} \beta'^{2}_{0}$.
The value of $\sigma$ is introduced to describe the intensity of turbulent motion
and the value of $p$ is used to describe the proportion of active cells.
Following each $\Delta t_{\rm turb}$ interval,
we resample the four-velocity components for all turbulent cells.
By incorporating the bulk motion properties,
i.e., Equations~(\ref{Eq.fireball-gamma})-(\ref{Eq.fireball-rho}),
one can obtain the overall characteristics of the turbulent jet in the lab frame of the burst.

In our phenomenological framework,
different physical mechanisms driving turbulence are distinguished solely by their characteristic values of $\sigma$ and $p$.
For example, 
magnetic reconnection would eject high-velocity fluids and occur sporadically in space.
This scenario can be described by adopting a large value of $\sigma$ and a small value of $p$ (tagged as ``Type-B'' in this paper).
Magnetic reconnection would also drive a weak turbulent motion in the bulk region, 
which can be described by adopting a low value of $\sigma$ and a high value of $p$ (tagged as ``Type-A'').
\cite{Lazarian-2019ApJ...882..184L} highlights the crucial role of turbulence in enhancing reconnection efficiency in GRBs.
The driven turbulence in turn triggers more magnetic reconnection,
which overlies the turbulent cells of Type-A turbulence.
Such turbulence is tagged as ``Type-A+Type-B''.
Typically, turbulence exhibits energy cascades across different scales (\citealp{Schekochihin-2009ApJS..182..310S}),
of which the dynamic of turbulent eddies is similar to
the Type-A+Type-B turbulence.
Then, we neglected the cascade behavior of turbulence.

%%%%%%%%%%%%%%%%%%%%%%%%%%%%%%%
\subsection{Basic Settings in Our Simulations}
\textit{Simulation Settings\;}
In our simulations,
unless otherwise specified,
the fireball parameters are set as follows:
$L_{\rm j} = 10^{52}\, \mathrm{erg\, s^{-1}}$,
$\eta = 100$,
and $R_{0} = 10^{8}\, \mathrm{cm}$.\footnote{The parameters of $L_{\rm j}$, $\eta$, and $R_{\rm 0}$ determine the temperature of the jet at the injection radius and thus affect the injected photons' energy.
In simulations with different values of $L_{\rm j}$, $\eta$, and $R_{\rm 0}$, the radiation spectrum may be shifted globally,
and thus the turbulence-induced effects on the radiation spectrum of the photosphere remain unaffected.
Then, a jet with $L_{\rm j} = 10^{52}\, \mathrm{erg\, s^{-1}}$, $\eta = 100$, and $R_{0} = 10^{8}\, \mathrm{cm}$ is adopted as an example.
}
Seed photons are sampled from a Planck spectrum,
and injected radius $R_{\rm inj}$ is set at $\tau = 30$ with optical depth $\tau (R)=L_{\rm j} \sigma_{\rm T} /( 8 \pi m_{\rm p} c^{3} \beta \eta^{3} R)$.
In addition, the photon scatter is ceased at the radius of $\tau =0.1$.
Unless otherwise specified, the turbulence in the jet is set in the region with $\tau\in[0.1, 30]$.
The polar angle of the injection region is constrained to $\theta \leqslant 3^{\circ}$,
where $\theta=0$ is set at the axis of the jet.
The total number of injected seed photons exceeds $10^{5}$.
The scale factor $\chi$ for the turbulence cell is set to $10$.

\textit{Observations Production\;}
To generate mock observational data,
we use ProcessMCRaT code (\citealp{Lazzati-2016ApJ...829...76L,Parsotan-2018ApJ...853....8P,Parsotan-2018ApJ...869..103P,Parsotan-2020ApJ...896..139P}).
The observer is positioned at a distance of $5 \times 10^{14}\,\mathrm{cm}$ from the central engine,
with $1^{\circ}$ deviation from the jet axis as the viewing angle,
and accept photons within $3^{\circ}$ range along the light of sight\footnote{ 
In our model, the jet is unstructured with properties uniform in polar angle, and the turbulence statistics are isotropic.
Then, the observed emission is not affected by the choice of viewing angle and angular acceptance if the viewing angle is set along the direction of jet flow.
If the viewing angle is beyond the direction of the jet flow, the observed emission may be different,
e.g., the flux would be very low.
The main focus of this work is the effect of turbulence on the radiation spectrum, and thus the situation by setting the viewing angle beyond the direction of the jet flow is not studied.}.
Due to the Doppler shift and the geometric effect,
the observed photospheric emission should be described by a multi-color blackbody (\citealp{Pe'er-2008ApJ...682..463P}).
Specifically, 
the multi-color blackbody emission with a cutoff power-law distribution temperature (CPL-mBB,
\citealp{2022ApJ...934L..22D}), 
i.e., $\hat{f}(T)\propto (T/T_{\rm c})^q\exp[-(T/T_{\rm c})^s]$,
can well describe the photospheric emission.
Based on the MCRaT for a photosphere free of turbulence,
the obtained radiation spectrum can be well described by a CPL-mBB with $q = 2.8$ and $s = 0.9$\footnote{Based on the numeric calculations, \cite{2022ApJ...934L..22D} finds a CPL-mBB with $q = 3$ and $s = 1.2$ can well describe the photospheric emission. Such a radiation spectrum is narrower compared to the CPL-mBB with $q = 2.8$ and $s = 0.9$. This is due to the fact that photon scattering is set to cease at the radius of $\tau=1$ in \cite{2022ApJ...934L..22D}.}.
The CPL-mBB spectrum model is used to fit the radiation spectra obtained in this paper.

\section{Result}\label{Sec:Result}

\subsection{Spectral Broadening and High-energy Photon Formation}
\begin{figure}
\centering
\includegraphics[width=0.45\textwidth]{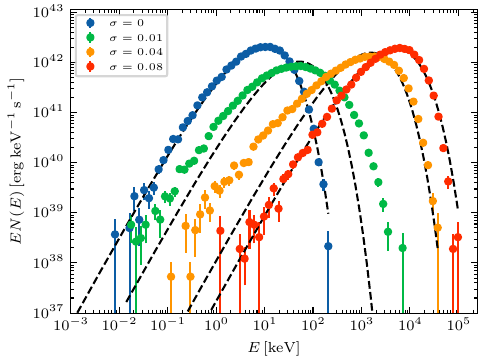}
\caption{Radiation spectra of the Type-A turbulent photosphere for different $\sigma$ values (fixed $p = 1$).
Black dashed lines show the corresponding spectral fitting based on the CPL-mBB model with $s=0.9$ and $q=2.8$.
}\label{Fig_Fnu_sigma0}
\end{figure}

\begin{figure}
\centering
\includegraphics[width=0.45\textwidth]{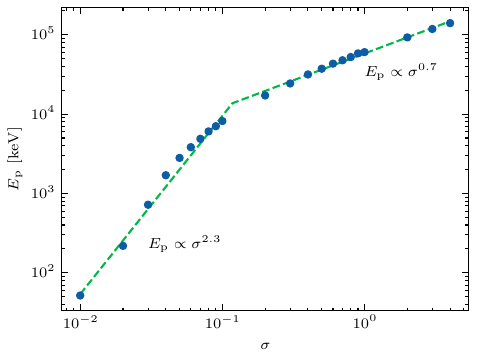}
\caption{Dependence of $E_{\rm p}$ on $\sigma$ for Type-A turbulence ($p=1$),
where the green dashed line indicates the best-fit with broken power law for $E_{\rm p} \text{--} \sigma$ relation.
}\label{Fig_sigma0_peak}
\end{figure}

We first study the impact of turbulent motion intensity $\sigma$ on the photospheric emission under Type-A turbulence with fixed $p=1$.
Figure~\ref{Fig_Fnu_sigma0} shows the radiation spectra of the turbulent photosphere with different values of $\sigma$.
One can find that the radiation spectra are shifted to the high-energy regime by increasing the value of $\sigma$.
This aligns with theoretical predictions that turbulence accelerates photons to higher energies via random scattering (\citealp{Murase-2012ApJ...746..164M}).
We fit the spectra using the CPL-mBB model with $q = 2.8$ and $s = 0.9$
and the results are indicated by black dashed lines in this figure.
It is notable that
the radiation spectra for the cases with $\sigma = 0.01$ or $\sigma=0.04$
are broader than the non-dissipative photospheric spectrum (i.e., the CPL-mBB model with $q = 2.8$ and $s = 0.9$).
However, the radiation spectra for the cases with very high $\sigma$, e.g., $\sigma=0.08$,
exhibit only a shift, with no change in spectral morphology.
The reason for this behavior will be discussed in detail in Section~\ref{Sec_Reason}.
Based on the fitting results with the CPL-mBB,
the spectral peak energy $E_{\rm p}$ in the $\nu \text{--} F_{\nu}$ radiation spectrum for different values of $\sigma$ is plotted in Figure~\ref{Fig_sigma0_peak}.
The relation of $E_{\rm p} \text{--} \sigma$ can be described with a broken power law,
where $E_{\rm p} \propto \sigma^{2.3}$ for $\sigma < 0.1$ and $E_{\rm p} \propto \sigma^{0.7}$ for $\sigma > 0.1$.

\begin{figure}
\centering
\includegraphics[width=0.45\textwidth]{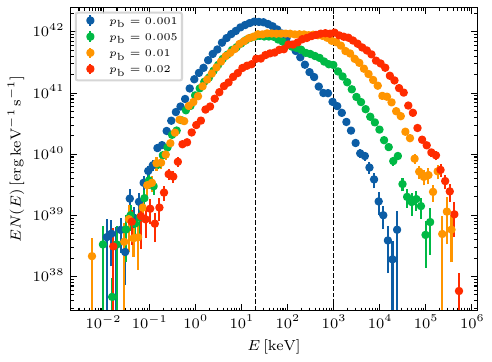}
\caption{Radiation spectra for Type-A+Type-B turbulence by varying $p_{\rm b}$,
where $(\sigma_{\rm a}, p_{\rm a})=(0.005, 1)$ and $(\sigma_{\rm b}, p_{\rm b})=(5, p_{\rm b})$ are set.
The left and right black dashed lines mark energies $E_{\rm 1} = 20\,{\rm keV}$ and $E_{\rm 2} = 1000\,{\rm keV}$, respectively.
}\label{Fig_Fnu_freq}
\end{figure}

\begin{figure}
\centering
\includegraphics[width=0.45\textwidth]{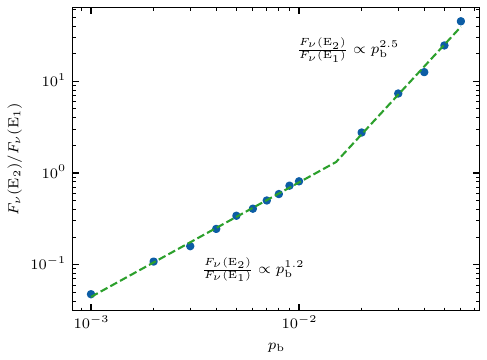}
\caption{Dependence of the flux density ratio $F_{\nu}(E_{\rm 2}) / F_{\nu}(E_{\rm 1})$ on $p_{\rm b}$,
with fixed $(\sigma_{\rm a},p_{\rm a}) = (0.005,1)$ and $(\sigma_{\rm b},p_{\rm b}) = (5,p_{\rm b})$.
The green dashed line indicates the best-fit broken power law to the $F_{\nu}(E_{\rm 2}) / F_{\nu}(E_{\rm 1}) \text{--} p_{\rm b}$ relation.}
\label{Fig_prob}
\end{figure}

\begin{figure}
\centering
\includegraphics[width=0.45\textwidth]{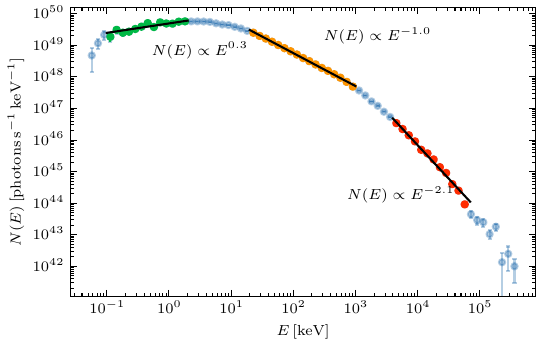}
\caption{Photon spectrum from MCRaT simulation with $\sigma_{\rm a}=0.005$, $p_{\rm a}=1$, $\sigma_{\rm b}=5$, and $p_{\rm b}=0.01$.
Black solid lines mark the best-fit power-law functions to three spectral segments: $N(E) \propto E^{0.3}$ (low-energy regime),
$N(E) \propto E^{-1.0}$ (mid-energy regime),
and $N(E) \propto E^{-2.1}$ (high-energy regime).
}\label{Fig_NE}
\end{figure}

We overlie cells of magnetic reconnection on a weak turbulent bulk region,
i.e., Type-A+Type-B turbulence.
For Type-A turbulence, we adopt $(\sigma, p)=(\sigma_{\rm a}, p_{\rm a})=(0.005, 1)$;
while for Type-B turbulence, we set $(\sigma, p)=(\sigma_{\rm b}, p_{\rm b})$ with $\sigma_{\rm b} = 5$ and $p_{\rm b}$ varying as specified.
Figure~\ref{Fig_Fnu_freq} presents radiation spectra for Type-A+Type-B turbulence at different values of $p_{\rm b}$.
It can be found that the high-energy regime is progressively intensified by increasing $p_{\rm b}$.
This result reveals that Type-B turbulence generates an additional component in the high-energy regime.
Moreover, this component is intensified by increasing $p_{\rm b}$.
Based on the case with $p_{\rm b} = 0.02$,
one can find that the additional component peaks at $E_{\rm 2} \sim 10^{3}\,\mathrm{keV}$,
which is distinct from the low-energy component peaking at $E_{\rm 1} \sim 20\,\mathrm{keV}$.
Figure~\ref{Fig_prob} shows the dependence of the flux ratio $F_{\nu}(E_{\rm 2})/F_{\nu}(E_{\rm 1})$ on the value of $p_{\rm b}$.
The relation follows a broken power law:
$F_{\nu}(E_{\rm 2})/F_{\nu}(E_{\rm 1}) \propto p_{\rm b}^{1.2}$ for $p_{\rm b}<10^{-1.8}$ and $\propto p_{\rm b}^{2.5}$ for $p_{\rm b}>10^{-1.8}$.
It is worth pointing out that the value of $F_{\nu}(E_{\rm 2})/F_{\nu}(E_{\rm 1})$ directly affects the spectral morphology of the photospheric emission.
For example, the photon spectral index $\alpha_{12}$ in the energy range of $E\in [E_1, E_2]$
can be estimated with $\alpha_{12}\sim \log [{F_\nu }({E_{\rm{2}}})/{F_\nu }({E_{\rm{1}}})]/\log [{E_{\rm{2}}}/{E_{\rm{1}}}]$.
For the case with $p_{\rm b} = 0.01$,
one can obtain $F_{\nu}(E_{\rm 2})/F_{\nu}(E_{\rm 1}) \sim 1$ and thus $\alpha_{12}\sim 0$.
As shown in Figure~\ref{Fig_NE},
the radiation spectrum for $p_{\rm b} = 0.01$ can be decomposed into three segments:
$N(E) \propto E^{0.3}$ for $E\in [0.1, 2]\,\mathrm{keV}$ (green data points),
$E^{-1}$ for $E\in [20, 1000]\,\mathrm{keV}$ (orange data points),
and $E^{-2.1}$ for $E\in [4000, 70000]\,\mathrm{keV}$ (red data points).
This is consistent with our above analysis.
The morphology of the middle segment (i.e., the orange data points) matches the low-energy regime of typical GRB spectra,
and the high-energy segment (i.e., the red data points) resembles the high-energy regime of typical GRB spectra.

\subsection{Turbulent Region Effect}\label{Sec_Reason}
\begin{figure}
\centering
\includegraphics[width=0.45\textwidth]{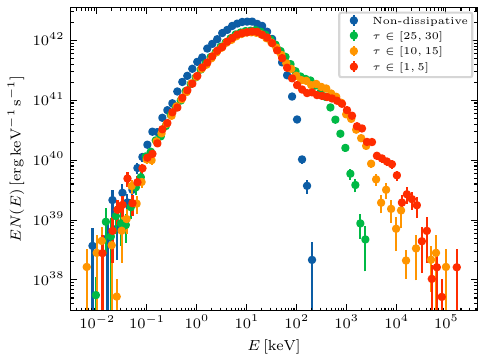}
\caption{Photospheric emission with turbulence implemented in different regions with $\tau\in[25,30]$ (green), $\tau \in [10,15]$ (orange), or $\tau \in [1,5]$ (red).
Type-A+Type-B turbulence with $(\sigma_{\rm a},p_{\rm a})=(0.005, 1)$ and $(\sigma_{\rm b},p_{\rm b})=(5.0,0.01)$ are adopted.
For comparison, the photospheric emission free of turbulence is also plotted with blue points.
}\label{Fig_Fnu_depth}
\end{figure}

The appearance of the high-energy peak component arises from incomplete thermalization of photons, which are scattered by turbulence.
Since the thermalization of photons is related to the optical depth $\tau$,
we study the photospheric emission by setting turbulence in different regions,
as shown in Figure~\ref{Fig_Fnu_depth}.
Here, 
we configure Type-A+Type-B turbulence with $(\sigma_{\rm a},p_{\rm a})=(0.005, 1)$ and $(\sigma_{\rm b},p_{\rm b})=(5.0,0.01)$.
Turbulence is implemented in the region across three optical depth ranges: $\tau\in[25, 30]$ (green),
$[10,15]$ (orange),
and $[1, 5]$ (red).
For comparison, 
we also plot turbulence-free photospheric emission (blue points).
One can find that the high-energy regime is intensified for a turbulent photosphere.
Furthermore, such intensification weakens as the turbulent region shifts to the inner region of the jet (corresponding to increasing $\tau$).
Since the thermalization of the photons is strengthened as $\tau$ increases, the distinct spectral signatures in the high-energy regime reflect varying degrees of thermalization efficiency.
Notably, when turbulence is implemented in the $\tau\in[1, 5]$ region (red points), a shoulder emerges in the high-energy regime of the spectrum.
This radiation spectrum profile resembles that of GRB~171205A's prompt emission.

\begin{figure}
\centering
\includegraphics[width=0.45\textwidth]{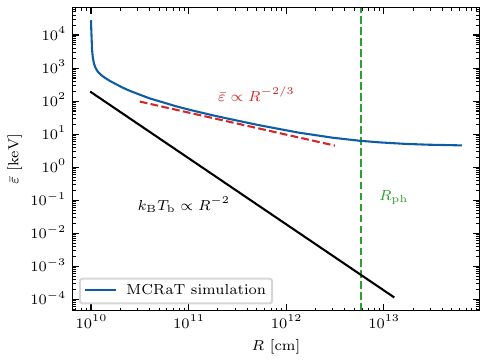}
\caption{Demonstration of the thermalization for photons: $\bar{\varepsilon} \text{--} R$ (blue line) vs. $k_{\rm B}T_{\rm b}-R$ (black line),
where a coasting jet with $k_{\rm B}T_{\rm b} \propto R^{-2}$ (black line) is adopted. 
The red dashed line depicts the theoretical prediction $\bar{\varepsilon} \propto R^{-2/3}$ (\citealp{Pe'er-2008ApJ...682..463P})
and the vertical green dashed line indicates the radius of $\tau = 1$.}\label{Fig_Ep_R}
\end{figure}

\cite{Pe'er-2008ApJ...682..463P} shows that the co-moving photon energy of a coasting jet
decreases with radius as $\propto R^{-2/3}$, 
owing to the slight misalignment of the scattering electrons' velocity vectors. 
Moreover, 
this mechanism dominates photon cooling.
We thus unfold electron thermalization effects by simulating a coasting jet with $T'_{\rm b} \propto R^{-2}$ for its electrons' temperature.
Results are shown in Figure~{\ref{Fig_Ep_R}},
where seed photons (with initial photon temperature set to 50 times the local electron temperature) are injected at the radius of $10^{10}\,{\rm cm}$.
The evolution of average photon energy $\bar{\varepsilon}$ versus radius $R$ is plotted as a blue solid line, while $k_{\rm B} T_{\rm b}$ versus $R$ appears as a black solid line, 
where $k_{\rm B}$ is the Boltzmann constant.
The strength of thermalization can be inferred by comparing the $\bar{\varepsilon} \text{--} R$ and $k_{\rm B} T_{\rm b} \text{--} R$ relations.
Strong photon thermalization is inferred if the $\bar{\varepsilon} \text{--} R$ curve matches the $k_{\rm B} T_{\rm b} \text{--} R$ curve,
while weak thermalization is indicated if deviations between the two curves occur.
Figure~{\ref{Fig_Ep_R}} reveals strong thermalization at the injection radius,
establishing thermal equilibrium between photons and electrons.
As the jet expands, the thermalization weakens
and the co-moving photon energy follows $\bar{\varepsilon} \propto R^{-2/3}$,
consistent with the theoretical expectation reported in \cite{Pe'er-2008ApJ...682..463P}.
Notably, an extensive region below the photosphere exhibits weak thermalization of photons by electrons.

\subsection{Reconcile with Band Function}
Most Fermi GRBs exhibit prompt emission spectra well described by a Band function with characteristic parameters $\alpha \sim -0.8$, $\beta \sim -2.3$, and $E_{\rm 0} \sim 150\,\mathrm{keV}$.
In this section, 
we present two realizations of the turbulent photosphere model,
whose emission is reconciled with the majority of Fermi GRBs.

As shown in Figures~\ref{Fig_Fnu_sigma0}, \ref{Fig_Fnu_freq}, and \ref{Fig_Fnu_depth}, thermal characteristics persist in the low-energy regime.
To reconcile the model with Fermi GRB observations, the spectral peak of a turbulence-free fireball must be below the \emph{Fermi}-GBM detection threshold (i.e., $8\,\mathrm{keV}$).
The spectral peak of the fireball is related to the observed temperature of the photosphere,
i.e., $T_{\rm ph} =(T_{\rm 0}/2)(R_{\rm ph} / R_{\rm s})^{-{2}/{3}}$,
where $R_{\rm ph} = L_{\rm j} \sigma_{\rm T} / (8 \pi m_{\rm p} c^{3} \beta \eta^{3}) \approx L_{\rm j} \sigma_{\rm T} / (8 \pi m_{\rm p} c^{3} \eta^{3})$ and $\beta=\sqrt{1-1/\Gamma_{\rm b}^2}$.
The $\nu \text{--} \nu F_{\nu}$ spectral peak photon energy is around $\xi k_{\rm B} T_{\rm ph} \approx 24.7\,\mathrm{keV}\, \eta_{2}^{8/3} L_{\rm j, 52}^{-5/12}$ with $\xi = 9.3$ based on MCRaT simulations.
Then, we take a jet with $\eta=50$ and $L_{\rm j}=10^{52}\,{\rm erg}\,{\rm s}^{-1}$ as an example.

\begin{figure}
\centering
\subfigure[]{
\includegraphics[width=0.45\textwidth]{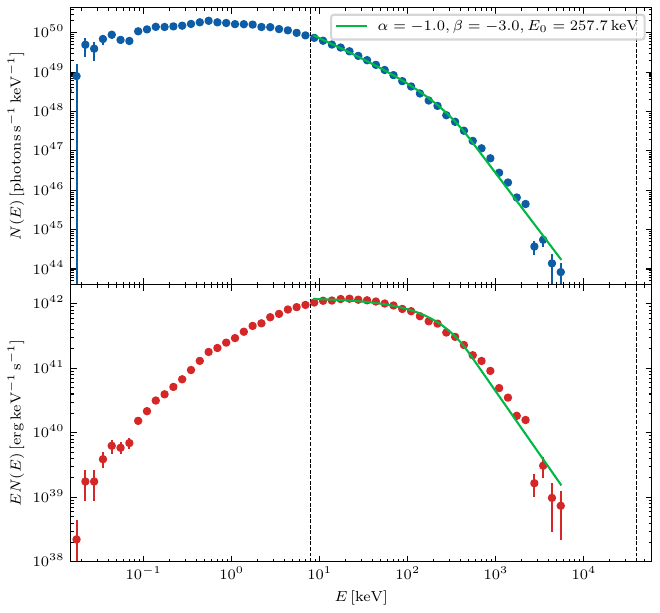}
\label{Fig_result_a}
}
\subfigure[]{
\includegraphics[width=0.45\textwidth]{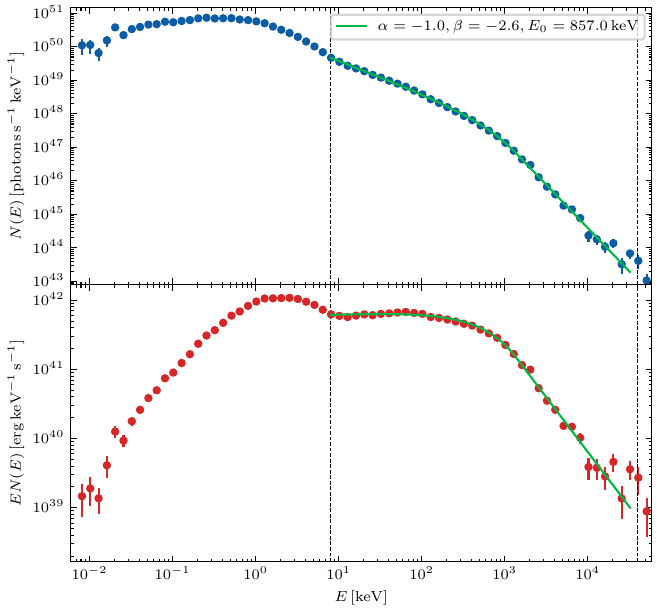}
\label{Fig_result_b}
}
\caption{Demonstration of two cases to reproduce the Band-like radiation spectrum:
(a) Type-A turbulence with $(\sigma,p)=(0.04,0.45)$ and turbulence implemented in the region of $\tau\in[0.1,30]$,
(b) Type-B turbulence with $(\sigma,p)=(7,0.05)$ and turbulence implemented in the region of $\tau\in[10, 15]$.
The blue and red data points show the photon spectrum $N(E)$ and
the energy spectrum $E N(E)$, respectively. 
The green solid line displays the best-fit Band function modeled over the \emph{Fermi}-GBM characteristic energy range $8\,\mathrm{keV}$ to $40\,\mathrm{MeV}$.
}\label{Fig_result}
\end{figure}

Figure~\ref{Fig_result_a} shows photospheric emission with Type-A turbulence implemented at $\tau < 30$, with parameters $\sigma = 0.04$ and $p = 0.45$.
The modeled turbulence produces a Band-like spectrum in the energy range of [$8\,\mathrm{keV}$, $40\,\mathrm{MeV}$], with spectral parameters $\alpha = -1.0$,
$\beta = -3.0$, and  $E_{\rm 0} = 257.7\,\mathrm{keV}$.
Figure~\ref{Fig_result_b} shows photospheric emission with Type-B turbulence implemented in the region $10 < \tau < 15$, using parameters $\sigma = 7$ and $p = 0.05$.
The scattered photons together with the thermal emission produce two segments of the radiation spectrum.
The Band function fit to the \emph{Fermi}-GBM's characteristic energy range yields $\alpha=-1.0$, $\beta=-2.3$, and $E_{\rm 0}=856.9\,\mathrm{keV}$.
Although the elevated $E_{\rm 0}$ exceeds typical values,
such high-energy peaks are occasionally observed in exceptional GRBs.
Figure~\ref{Fig_result} demonstrates that the turbulent photospheres can reproduce both typical and extreme GRB spectral features.

\section{Conclusions and Discussion}\label{Sec:Discussion}
This paper is focused on the effect of turbulence on the radiation spectrum of the photospheric emission.
Our simulations based on the MCRaT code demonstrate that
the turbulence below the photosphere plays a significant role in shaping the photospheric emission.
The effect of the turbulence on the photospheric emission critically
depends on the spatial distribution and velocity of turbulent eddies.
If the turbulence is well developed under the photosphere of a jet,
low-velocity turbulence can significantly intensify the photospheric emission in the high-energy regime
by scattering soft photons to high-energies.
A spectral bump may appear in the high-energy regime.
Its peak photon energy increases with turbulent velocity, while its flux density $F_{\nu}$ rises with increasing spatial occupancy ratio of eddies.

The weak coupling between photons and electrons is required in order to
preserve the distortion of the radiation field by the turbulence.
We find that a wide region below the photosphere appears
to exhibit weak coupling between electrons and photons.
If the proportion of neutrons in the jet is significant,
electrons and photons decouple more easily (e.g., \citealp{2024ApJ...965....8W}).
We also demonstrate that the observed radiation spectra of \emph{Fermi}-GBM on GRBs can be reproduced
if a turbulent fireball with $k_{\rm B}T_{\rm ph} \lesssim1\,{\rm keV}$ is adopted,
where $T_{\rm ph}$ is the observed temperature of its photosphere.
If the contribution of synchrotron radiation from electrons is considered,
e.g., \cite{2021ApJ...922..257P},
the $k_{\rm B}T_{\rm ph} \lesssim1\,{\rm keV}$ requirement for the turbulent fireball can be relaxed.

\begin{acknowledgments}
We acknowledge Dr. Parsotan for the publicly available MCRaT code. 
We thank the anonymous referee of this work for useful comments and suggestions that improved the paper.
This work is supported by the National Natural Science Foundation of China (grant Nos. 12273005, 12494575, and 12133003),
the National Key R\&D Program of China (grant No. 2023YFE0117200 and 2024YFA1611700),
the special funding for Guangxi Bagui Youth Scholars,
and the Guangxi Talent Program (``Highland of Innovation Talents'').
\end{acknowledgments}

\software{MCRaT (\cite{Lazzati-2016ApJ...829...76L,Parsotan-2018ApJ...853....8P,Parsotan-2018ApJ...869..103P,Parsotan-2020ApJ...896..139P}),
ProcessMCRaT (\citealp{Lazzati-2016ApJ...829...76L,Parsotan-2018ApJ...853....8P,Parsotan-2018ApJ...869..103P,Parsotan-2020ApJ...896..139P}),
matplotlib (\citealp{Hunter:2007}),
numpy (\citealp{harris2020array}).
}

%%%%%%%%%%%%%%%%%%%%%%%%%%%%%%%%%%%%
\bibliography{bibliography}

\clearpage

	%\begin{thebibliography}{}
		
		%\bibliography{bibliography}

	%\end{thebibliography}
	
\end{document}